\documentclass[12pt,a4paper]{revtex4}
\usepackage{epsfig}                           
\begin{document}
\title{ Stochastic resonance and heat fluctuations in a driven double-well system}
\author{Mamata Sahoo$^{1}$, Shantu Saikia$^{2,3}$, Mangal C. Mahato$^{3}$ and A. M. Jayannavar$^{1}$}
\address{$^{1}${Institute of Physics,~~Sachivalaya Marg,~~ Bhubaneswar-751005, India}}
\address{$^{2}$ {St.Anthony's College, Shillong-793003, India}}
\address{$^{3}${Department of Physics, North-Eastern Hill University,
 Shillong-793022, India}}
\begin{abstract}
Abstract:~~We study a periodically driven (symmetric as well as asymmetric)double-well potential system at finite temperature.~~We show that mean heat loss by the system to the environment~~(bath) per period of the applied field is a good quantifier of stochastic resonance.~~~It is found that the heat fluctuations over a single period are always larger than the work fluctuations.~~~The observed distributions of work and heat exhibit pronounced asymmetry near resonance.~~~The heat losses over a large number of periods satisfies the conventional steady-state fluctuation theorem,~~though different relation exists for this quantity.
\end{abstract}
\maketitle
\maketitle

\vspace{3.5cm}
\noindent
\newcommand{\nwc}{\newcommand}
\nwc{\bdm}{\begin{displaymath}}
\nwc{\edm}{\end{displaymath}}

Corresponding Author:  A.M. Jayannavar\\
Email address       :  jayan@iopb.res.in 
\newpage
\section{Introduction}
Stochastic Resonance (SR) was discovered barely about two and half decades
ago, yet it has proved to be very useful in explaining many phenomena in 
natural sciences[1-3]. SR refers to an enhanced response of a nonlinear system 
to a subthreshold periodic input signal in the presence of noise of  optimum 
strength. Here, noise plays a constructive role of pumping power in a 
particular mode, that is in consonance with the applied field, at the cost of
the entire spectrum of modes present in it. SR, so defined, leaves a lot of 
liberty as to what is the physical quantity that is to be observed which 
should show a maximum as a function of noise strength[4-23]. In other words, 
no unique quantifier of SR is specified. Also, in order that SR be a bonafide
resonance the quantifier must show maximum as a function of frequency of the 
applied field as well. For instance, in a double-well system, hysteresis loop
area, input energy or work done on the system in a period of the driving 
field and area under the first peak in the residence time (in a well) 
distribution are used to characterize SR as a bonafide resonance[4-17,19-22].

In the present work, motivated by recently discovered fluctuation theorems,
we show that in an overdamped bistable system input energy per period as well
as the energy absorbed per period by the system from the bath, i.e, the heat,
can be used as quantifiers to study SR. Also, it is found that the relative 
variance of both the quantities exhibit minimum at resonance; that is, 
whenever input energy and heat show maximum as a function of noise strength 
(as also frequency), their respective relative fluctuations show minimum. 
This shows that at SR the system response exhibits greater degree of 
coherence. These fluctuations, however, are very large and often the physical
quantities in question become non-self-averaging. We study some of these 
aspects in the light of the fluctuation theorems in the following sections. 
The fluctuation theorems are of fundamental importance to nonequilibrium 
statistical mechanics[24-46]. The fluctuation theorems describe rigorous 
relations for properties of distribution functions of physical variables such
as work, heat, entropy production, etc., for systems far from equilibrium 
regimes where Einstein and Onsagar relations no longer hold. These theorems 
are expected to play an important role in determining thermodynamic 
constraints that can be imposed on the efficient operation of machines at  
nano scales. Some of these theorems have been verified experimentally[47-53].

\section{The Model}
We consider the motion of a particle in a double-well potential 
$V(x)=-\frac{a x^{2}}{2}+\frac{b x^{4}}{4}$ under the action of a weak 
external field $h(t)=A\sin(\omega t)$. The motion is described by the 
overdamped Langevin equation[44]
\begin{equation}
\gamma \frac{dx}{dt}=-\frac{\partial U(x)}{\partial x}+\xi(t) ,
\end{equation}
where $U(x)=V(x)-h(t)x$. The random forces satisfy 
$\langle \xi(t) \rangle =0$ and 
$\langle \xi(t)\xi(t^{'}) \rangle=2\gamma k_{B} T \delta(t-t^{'})$,
where $\gamma$ is the coefficient of friction, $T$ is the absolute 
temperature and $k_{B}$ is the Boltzmann constant. In the following we use a 
dimensionless form of equation(1), namely,
\begin{equation}
\frac{dx}{dt}=-\frac{\partial U(x)}{\partial x}+\xi(t),
\end{equation}
where $U(x)=-\frac{x^{2}}{2}+\frac{x^{4}}{4}-xh(t)$, and 
the external field $h(t)=A\sin(\omega t)$. Now, $\xi(t)$ satisfies 
$\langle \xi(t) \xi(t^{'}) \rangle=D \delta(t-t^{'})$, where $D=2 k_{B} T$. 
All the parameters are given in dimensionless units~~(in terms of 
$\gamma$, $a$ and $b$). We consider $A \ll 0.25$, so that the forcing 
amplitude is much smaller than the barrier height between the two wells.

Following the stochastic energetic formalism developed by Sekimoto[55], the 
work done by the external drive $h(t)$ on the system or the input energy per 
period (of time $\tau_{\omega}$) is defined as[21] 
\bdm
W_{p}= \int_0^{t_{0}+\tau_{\omega}} \frac{\partial U}{\partial t} dt
\edm
\begin{equation}
= -\int_0^{t_{0}+\tau_{\omega}} x(t) \frac{dh(t)}{dt} dt,
\end{equation} 
where $h(t)$ is the drive field which completes its period in time 
$\tau_{\omega}$. The completion of one period of $h(t)$, however, does not 
guarantee the system coming back to the same state as the starting one. In 
other words, $x(t+\tau_{\omega})$ need not be equal to $x(t)$ or 
$U(x,t+\tau_{\omega})$ may differ from $U(x,t)$. The work done over a period 
$W_{p}$ equals  change in the internal energy 
$\Delta U=U(x,t_{0}+\tau_{\omega})-U(x,t_{0})$ and heat $Q$ absorbed over a 
period (first law of thermodynamics), i.e, $W_{p}=\Delta U_{p}+Q_{p}$. Since 
$x(t)$ is stochastic, $W_{p}$, $\Delta U_{p}$ and $Q_{p}$ are not the same 
for different cycles(or periods) of $h(t)$. The averages are  evaluated from 
a single long trajectory $x(t)$ (eqn(3)). From the same calculations one can 
also obtain the probability distribution $P(W)$ and various moments of $W$.
Similarly, appealing to the first law of thermodynamics as stated above we 
can obtain $P(Q_{p})$ and $P(\Delta U_{p})$ and their moments, where the 
subscript p indicates evaluation of the physical quantities over one period 
of the field. Numerical simulation of our model was carried out by using 
Huen's method[56]. To calculate $W_{p}$ and $Q_{p}$ we first evolve the 
system and neglect initial transients. To get better statistics we calculate 
$W_{p}$, $Q_{p}$ for $10^{6}$ cycles. In some cases we evaluate $W$, 
$\Delta U$ and $Q$ over many periods, $n$, and calculate their averages, 
again, for $10^6$ such entities.
\section{Results and Discussions}
The internal energy being a state variable, average change in its value over 
a period $\Delta U_{p}$ is identically equal to  zero. Thus, in the time 
periodic asymptotic state averaged work done over the period 
$\langle W_{p} \rangle$ is dissipated in to heat $\langle Q_{p} \rangle$ by 
the system to the bath. Thus, $\langle Q_{p} \rangle$ can also be identified 
as hysteresis loop area. As has been reported earlier[19-22], 
$\langle W_{p} \rangle$, the input energy per period, shows a maximum as a 
function of $D$. Fig(1) shows that $\langle W_{p} \rangle$ and 
$\langle Q_{p} \rangle$ coincide, thus both the physical quantities show
SR. Hence, in this case input energy per period, the heat per period or the 
hysteresis loop area can equally well quantify stochastic resonance. However,
in this work we focus mostly on the fluctuation properties of these 
quantities.

The relative variances $R_{W}$ and $R_{Q}$ of both $W_{p}$ and $Q_{p}$ 
respectively show minimum (fig(2)) as a function of $D$.  It may be 
noted that even though $\langle W_{p} \rangle$ and $\langle Q_{p} \rangle$ are identical,~~  fluctuations in $W_{p}$ differ from the fluctuations in  $Q_{p}$.
 The relative variance of $Q_{p}$ is always larger than that of $W_{p}$ for all $D$. It is also noteworthy that the minimum value of the relative 
variance is  larger than one. However, the minimum becomes less than one if 
the averages are taken not over a single period of the field but over a 
larger(integral) number, $n>1$, of periods. Therefore, in order to obtain 
meaningful averages of these physical quantities in such driven systems one 
needs to study over time scales much larger than one period so that the 
averages are significantly larger than the deviations about them. Also, as 
$n$ becomes large, the differences between the relative variances of $W$ and 
$Q$ become insignificant(see inset of fig(2)). Importantly, in the system 
under study, this situation (mean $>$ dispersion) can be achieved by 
increasing the duration of averaging time(or the number of periods,~ $n$) more 
easily around the value of $D$ where SR occurs. The minimum of relative 
variance occurs just because the mean value is largest there and not because 
dispersions are smallest. However, as the number of periods $n$ is increased 
the mean value of heat dissipated over the $n$ periods 
$\langle Q_{np} \rangle \sim n$ for all $n$, whereas the dispersion 
$\sim \sqrt{n}$ for large $n$ so that the relative variance decreases with 
$n$ as $\frac{1}{\sqrt{n}}$ and one gets a range of $D$ where the averages 
become meaningful. We have observed numerically that $Q_{np}$ behaves as an 
independent variable only when evaluated over a larger number of cycles $n$ 
as compared to in case of $W_{np}$. For our present parameters  approximately
$Q_{np}$ is uncorrelated beyond $10$ periods, whereas $W_{np}$ is uncorrelated beyond $5$ periods.

In fig(3), we have plotted average heat dissipated 
$\langle Q_{p} \rangle$($=\langle W_{p} \rangle$) over a single period as a 
function of frequency. The values of physical parameters are given in the 
figure caption. The figure shows maximum as shown in earlier literature[21].
Thus $\langle Q_{p} \rangle $ acts as a quantifier of bonafide stochastic 
resonance. In the inset we give the corresponding relative variance of heat 
and work as a function of frequency. We observe that heat fluctuations are 
larger than work fluctuations at all frequencies. Near the resonance the 
relative variance shows a minimum. It may be noted that minimum relative 
variance of both quantities $W_{p}$ and $Q_{p}$ are larger than one(fig(2) and fig(3)). 

In fig(4), we plot the probability distribution of $W_{p}$ and $Q_{p}$ for 
various values of $D$. For low values of $D$ (e.g., $D=0.02$) $P(W_{p})$ is 
Gaussian  whereas $P(Q_{p})$ has a long exponential tail as in case of a 
system driven in a harmonic well and with almost no chance of a particle 
going over to the other well of the double-well potential. As $D$ is 
gradually increased rare passages to the other well becomes a possibility and
a very small peak appears at a finite positive value of $W_{p}$(or $Q_{p}$)
(e.g., at $D=0.04$). As $D$ is increased further, $P(W_{p})$ and $P(Q_{p})$ 
become multipeaked and the averages $\langle W_{p} \rangle$, 
$\langle Q_{p} \rangle$ shifts to their positive values. The distributions 
become most asymmetric at around $D=0.12$ (where SR occurs) and the asymmetry
reduces again at larger $D$, fig(4). When $D$ becomes large (e.g., $D=0.5$) 
the distribution becomes completely symmetric and at such high $D$ values the
presence of potential hump becomes ineffective to split the distribution into
two or more peaks. At very small and very large $D$ values $P(W_{p})$ is close to 
Gaussian and so does $P(Q_{p})$ but with a slow decaying exponential tail. In all
the graphs, the distribution of $P(Q_{p})$ ($P(W_{p})$) extend to negative values of 
$Q_{p}$ ($W_{p}$). Finite value for distribution in the negative side is 
necessary  to satisfy certain fluctuation theorems. Moreover, $P(Q_{p})$
has higher  weightage for large negative $Q_{p}$ than that of work $W_{p}$.

It is worth reemphasizing that $W$ and $Q$ behave as additive (or extrinsic) 
physical quantities with respect to the number of periods $n$ and hence 
$\langle W_{np} \rangle $ and $\langle Q_{np} \rangle $ increase 
in proportion to $n$ whereas $\Delta U$, in this case, is an intrinsic 
physical quantity and  $\frac{\Delta U}{n}  \rightarrow  0$ as 
$n \rightarrow \infty$. This indicates that the distributions $P(W_{np})$ and
$P(Q_{np})$ both have  identical characteristics as  $n \rightarrow \infty$.
Therefore, the difference between 
$(\frac{\sqrt{\langle W_{np}^{2} \rangle -\langle W_{np} \rangle ^{2}}}
{\langle W_{np} \rangle})$ and 
$(\frac{\sqrt{\langle Q_{np}^{2} \rangle -\langle Q_{np} \rangle ^{2}}}
{\langle Q_{np} \rangle})$ vanishes as $n \rightarrow \infty$. In the recent 
literature it is shown that  the distribution $P(W_{np})$ over a large number
of periods approaches a Gaussian. Also, if  one considers  $W_{p}$ over a 
single period by increasing the noise strength, $P(W_{p})$ approaches 
Gaussian and satisfies the steady state fluctuation theorem (SSFT). SSFT 
implies[26,34-36,44-46,51-53] the probability of physical quantity $x$ to 
satisfy the relation  $P(x)/P(-x) = \exp(\beta x)$, where $\beta$ is 
the inverse temperature and $x$ may be work, heat, etc. In fig(5), the 
evolution of $ P(Q_{np})$ is shown as $n$ is increased . As $n$ increases the 
contribution of negative $Q$ to the distribution decreases; besides, the 
distribution gradually becomes closer and closer to Gaussian. There is a 
contribution to $P(Q_{np})$ due to change in the internal energy $\Delta U$ 
which is supposed to dominate at very large $Q$ making the distribution 
exponential in the asymptotic regime[34,35,53]. However, it is not possible to 
detect this exponential tail in our simulations. For large $n$, $P(Q_{np})$  
approaches Gaussian(inset of fig(5)). The Gaussian fit of the graph almost 
overlaps and the calculated ratio, 
$\frac{\langle Q_{np}^{2} \rangle -\langle Q_{np} \rangle^{2}}{\frac{2}
{\beta} \langle Q_{np} \rangle}$ equals $0.99$ for $n=25$. This ratio is 
closer to one, a requirement for SSFT to hold where $P(Q)$ is Gaussian[22,44,45]. 
Fig(6) shows the plot of $ln(\frac{P(Q_{np})}{P(-Q_{np})})$ as a function of 
$\beta Q_{np}$ for various values of $n$. One can readily see that slope of 
$ln(\frac{P(Q_{np})}{P(-Q_{np})})$ approaches $1$ for 
$Q \ll \langle Q_{np} \rangle $ for large $n$. This is a statement of 
conventional steady state fluctuation theorem. As the number of periods $n$,
over which $Q_{np}$ is calculated, is increased, the conventional SSFT is 
satisfied  for $Q_{np}$ less than $\langle Q_{np} \rangle$ (e.g., for $n=25$,
SSFT is valid for $Q_{np}$ less than $0.4$, for $D=0.16$). There exists an 
alternative relation for heat fluctuation, namely, the extended heat 
fluctuation theorem[34,35]. Here, the distribution function obeys a different 
symmetry property for $Q \gg \langle Q_{np} \rangle$ for finite $n$. As 
$n \rightarrow \infty$, $\langle Q_{np} \rangle \rightarrow \infty$  in this 
limit, and hence conventional SSFT holds  which has been clarified earlier 
in linear systems[53]. 

It is further interesting to investigate effects associated with SR in an asymmetric double-well potential involving two hopping time scales instead of one as in the symmetric case.~~~We therefore,~~~consider a scaled asymmetric potential $V(x)=\frac{- x^{2}}{2}+\frac{x^{4}}{4}-cx$ driven by the external field $h(t)$.~~~~Fig(7) shows the average input energy $\langle W_{p} \rangle$ and average heat $\langle Q_{p} \rangle $ over a single period as a function of $D$ for various values of the asymmetric parameter $c$.~~~From this figure we find that the peak becomes broader and lower as $c$ is increased.~~~The peak shifts to larger values of noise intensities for higher $c$.~~~~In other words,~~the phenomenon of SR is not as pronounced[2] as in case of $c=0$(fig(2)).~~~It is because the synchronization between signal and particle hopping between the two well becomes weak because for $c \neq 0 $,~~~the mean time of passage for well $1$ to well $2$ is different from  the mean time of passage from  well $2$ to well $1$.~~~As a consequence the relative variances $R_{W}$ and $R_{Q}$ become larger as compared to in case of $c=0$(fig(2)) as shown in the inset of fig(7).

In fig(8(a)) and fig(8(b) we have plotted probability distribution $P(W_{p})$ and $P(Q_{p})$ over a single period for different values of asymmetry parameter $c$ for a fixed value of $D=0.12$, $A=0.1$ and $\omega =0.1$.~~~As asymmetry increases the probability for particle to remain in the lowest well enhances.~~Hence particle performs simple oscillation around most stable minima over a longer time before making transitions to the other well.~~~Hence Gaussian like peak near $W \approx 0$ or $Q \approx 0$ increases as $c$ increases.~~~The weight of $P(W_{p})$ for larger values of work(positive as well as negative ) decreases with increase in $c$.~~~However,~~~for $P(Q_{p})$, its magnitude at large positive and negative values of $Q_{p}$ increases as we increase asymmetry parameter.~~~This contrasting behavior can be attributed to the larger fluctuations of internal energy $\Delta U_{p}$ as one increases $c$.~~This we have verified
 separately.~~~Due to this contribution of $\Delta U_{p}$ for $Q_{p}$,~~nature of $P(W_{p})$ and $P(Q_{p})$ are qualitatively different.~~In all cases for fixed asymmetry $c$ fluctuation in heat are larger than fluctuation in work.
 
 In fig(9) and (10) evolution for $P(W_{np})$ and $P(Q_{np})$ respectively are plotted for various values of number of periods $n$.~~We clearly observe that as $n$ increases both the distributions tend to become  Gaussian distributions with the  fluctuation ratio $\frac {V}{(\frac{2}{\beta}\langle M \rangle )}=1$,~~between their variance $V$ and mean $\langle M \rangle$ as required to satisfy SSFT  as mentioned earlier.~~To satisfy SSFT for heat we have to take 
 larger number of periods as compared for work.~~Only in the large $n$ limit contribution to heat from internal energy becomes negligible.~~~In the insets of fig(9) and fig(10) we have shown a Gaussian fit(with fluctuation ratio equal to one),~~which agrees perfectly well with our numerical data.~~~Conclusions regarding validity of SSFT for asymmetric case for larger periods remain the same as for the symmetric case.

In summary, we find that SR shown by a particle moving in a double-well(symmetric) 
potential and driven by a weak periodic field can be characterized well by the heat~$\langle Q_{p}\rangle$ dissipated to the bath or the hysteresis loop area.~~~ It can equally well be characterized by the relative dispersion of $\langle W_{p}\rangle$ and $\langle Q_{p}\rangle$.~~~ At resonance relative dispersion shows a minimum as a function of both $D$ and $\omega$.~~~~ We also show that minimum relative variance can be made less than one by taking long time protocols of the applied field.~~~For long time protocols distribution  $P(Q_{np})$ satisfies conventional SSFT for $P(Q_{np})$ at  $Q_{np} \ll \langle Q_{np} \rangle$ for finite $n$[53].~~~We have also shown that SR gets weakened in the presence of asymmetric potential and as a consequence fluctuation in heat and work become larger.~~~SSFT too is satisfied for both work and heat,~~when they are calculated over large number of periods.
\section{Acknowledgements:}
AMJ and MCM thank BRNS, DAE,Govt. of India for partial financial support.
~~AMJ also thanks DST, India for financial support.~~ MCM acknowledges IOP, Bhubaneswar for hospitality.

\newpage
\section{Figure Captions}
Fig.1:~~The average input energy $\langle W_{p} \rangle$ and $\langle Q_{p} \rangle$ as a function of $D$ for $\omega=0.1$ and $A=0.1$.\\

Fig.2:~~~The relative variance $R_{W}$ and $R_{Q}$ over one period are plotted as a function of $D$.~~In the inset the relative variance $R_{W}$ and $R_{Q}$ over 25 periods are presented.~~The other parameters are same as in fig(1).\\

Fig.3:~~~The mean heat energy $\langle Q_{p} \rangle$ is plotted as a function of $\omega$ for $D=0.15$ and $A=0.1$.~~In the inset $R_{W}$ and $R_{Q}$ over one period are presented.\\

Fig.4:~~~The distribution $P(W_{P})$ and $P(Q_{P})$ over a single period for various values of $D$:~~$0.02(a)$,~$0.04(b)$,~$0.06(c)$,~$0.08(d)$,~$0.10(e)$,~$0.12(f)$,~$0.16(g)$ and $0.5(h)$.\\

Fig.5:~~~The evolution of $P(Q_{np})$ over different periods is presented.~~In the inset $P(Q_{np})$ over $25$ periods is plotted together with its Gaussian fit $f(Q)$.~Here $D=0.12$,~$A=0.1$ and $\omega=0.1$.\\

Fig.6:~~~The plot of $ln(P(Q_{np})/P(-Q_{np}))$ with temperature $\beta Q_{np}$  for different periods.~~Only the range of $Q_{np}$ is presented for which the curves are nearly linear.~~The parameters are same as that in fig(5) except that here $D=0.16$.\\

Fig.7:~~~The plot of $\langle W \rangle$ with temperature for various values of the asymmetry parameter,~$c$.~~Inset shows $R_{W}$ and $R_{Q}$ for $c=0.1$ .\\

Fig.8:~~~The distribution of $P(W_{P})$ and $P(Q_{P})$ over a single period  for $c=0.0$, $c=0.05$, $c=0.1$ and $c=0.15$.~Here $D=0.12$,$A=0.1$ and $\omega =0.1$.\\

Fig.9:~~~The evolution of $P(W_{np})$ over different periods for $c=0.1$.In the inset $P(W_{np})$ over  $25$  periods is plotted together with its  Gaussian fit $f(W)$.~~~Other parameters are same as fig(8).\\

Fig.10:~~~The evolution of $P(Q_{np})$ over different periods for 
$c=0.1$. In the inset $P(Q_{np})$ over  $25$  periods is plotted together with its Gaussian fit $f(Q)$.~Other parameters are same as fig(8).\\

\newpage
\begin{figure}[htp!]
\begin{center}
\input{epsf}
\includegraphics [width=6in,height=6.5in] {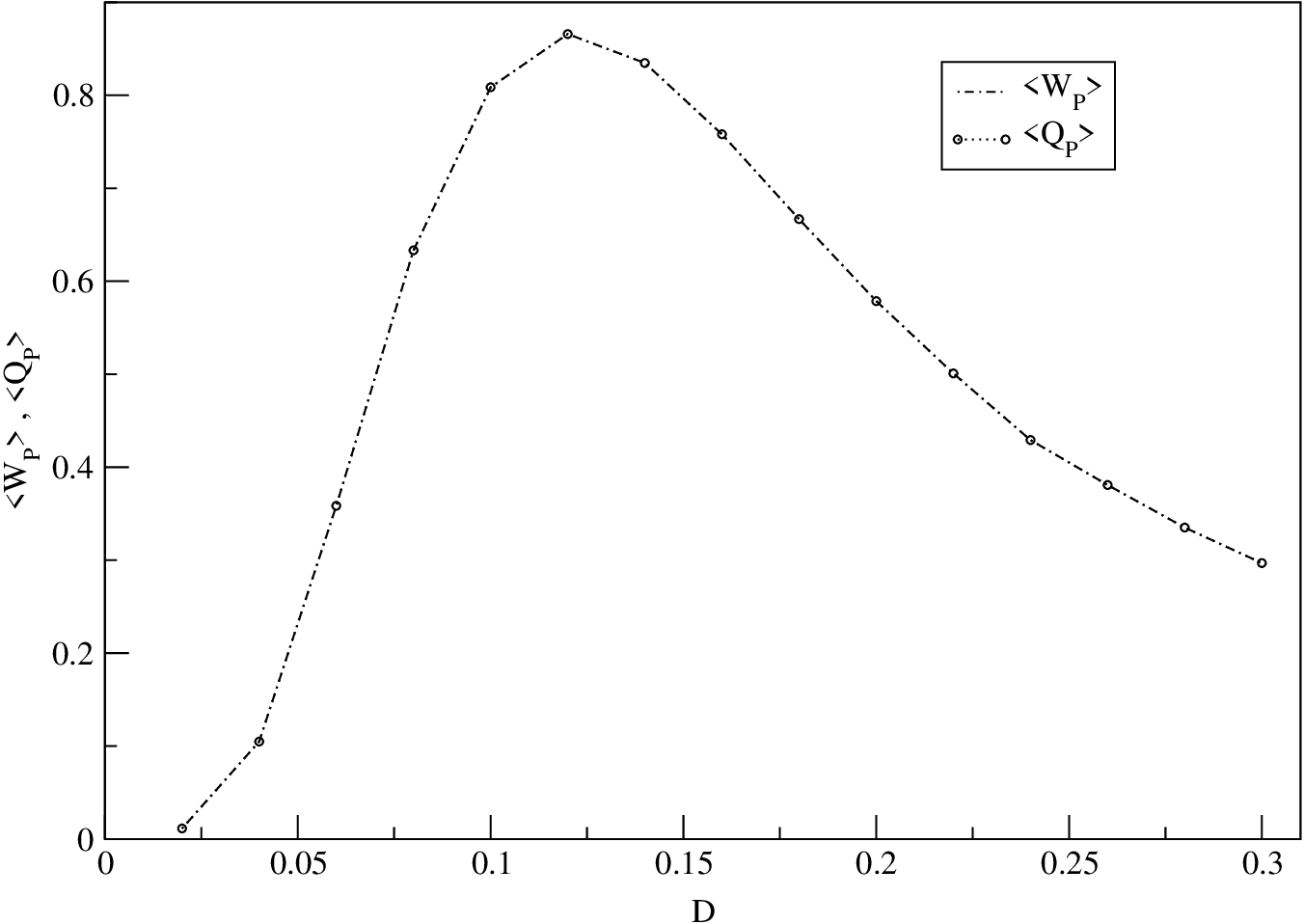}
\caption{}
 \end{center}
 \end{figure}

\newpage
\begin{figure}[htp!]
\begin{center}
\input{epsf}
\includegraphics [width=6in,height=6.5in] {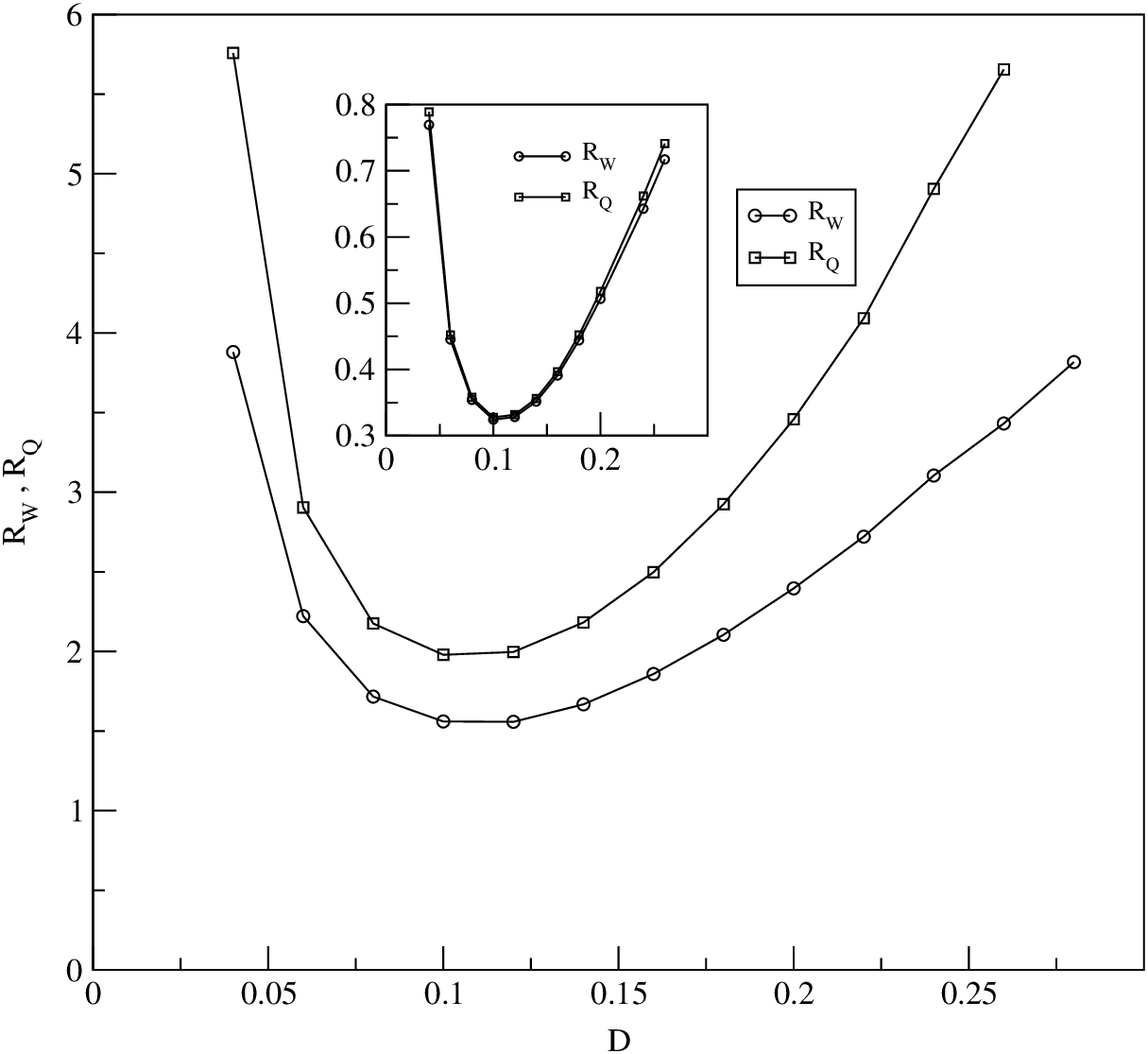}
\caption{} 
\end{center}
 \end{figure}
 
\newpage
\begin{figure}[htp!]
\begin{center}
\input{epsf}
\includegraphics [width=6in,height=6.5in] {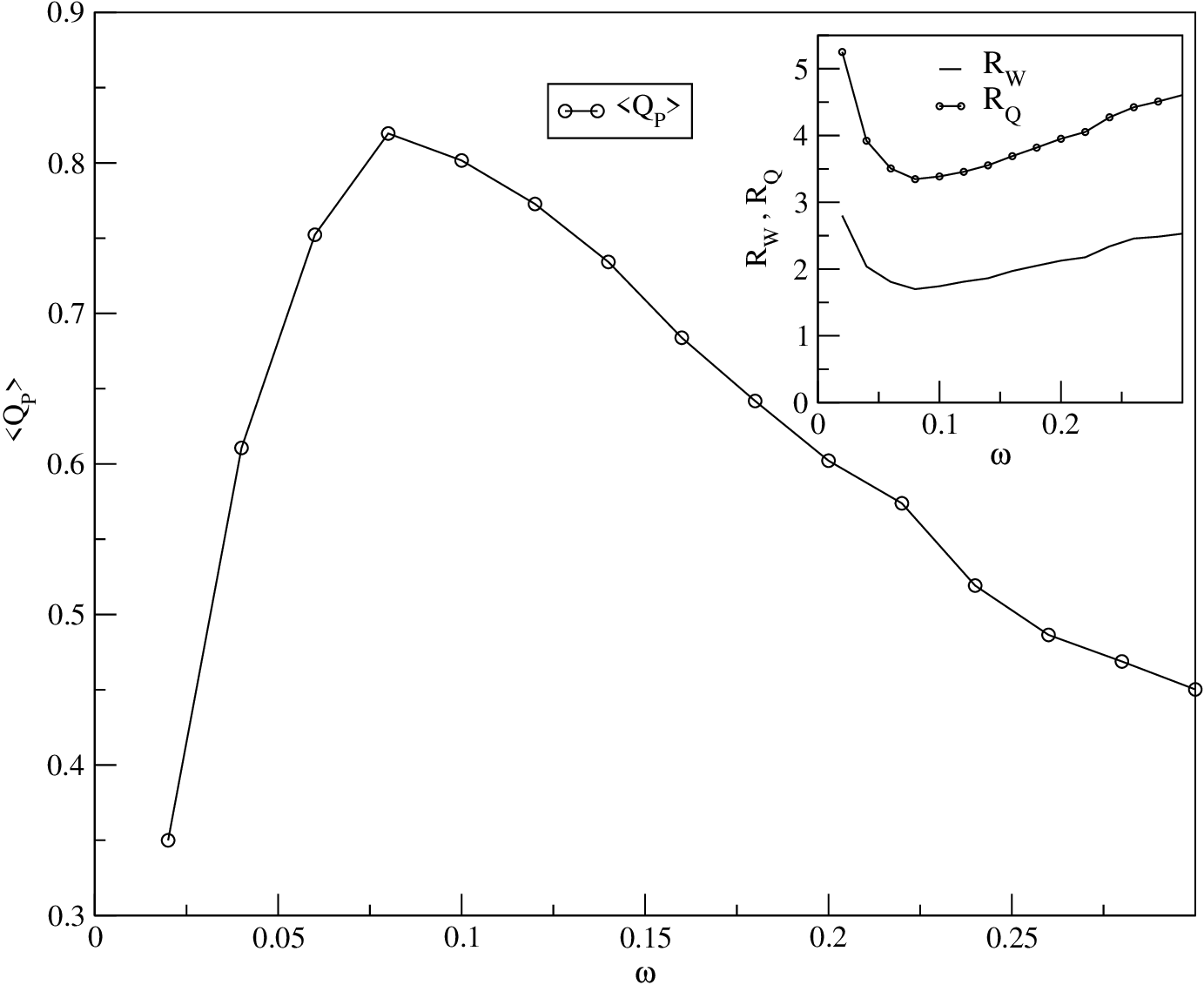}
\caption{} 
\end{center}
\end{figure}

\newpage
\begin{figure}[htp!]
\begin{center}
\input{epsf}
\includegraphics [width=6.1in,height=8in] {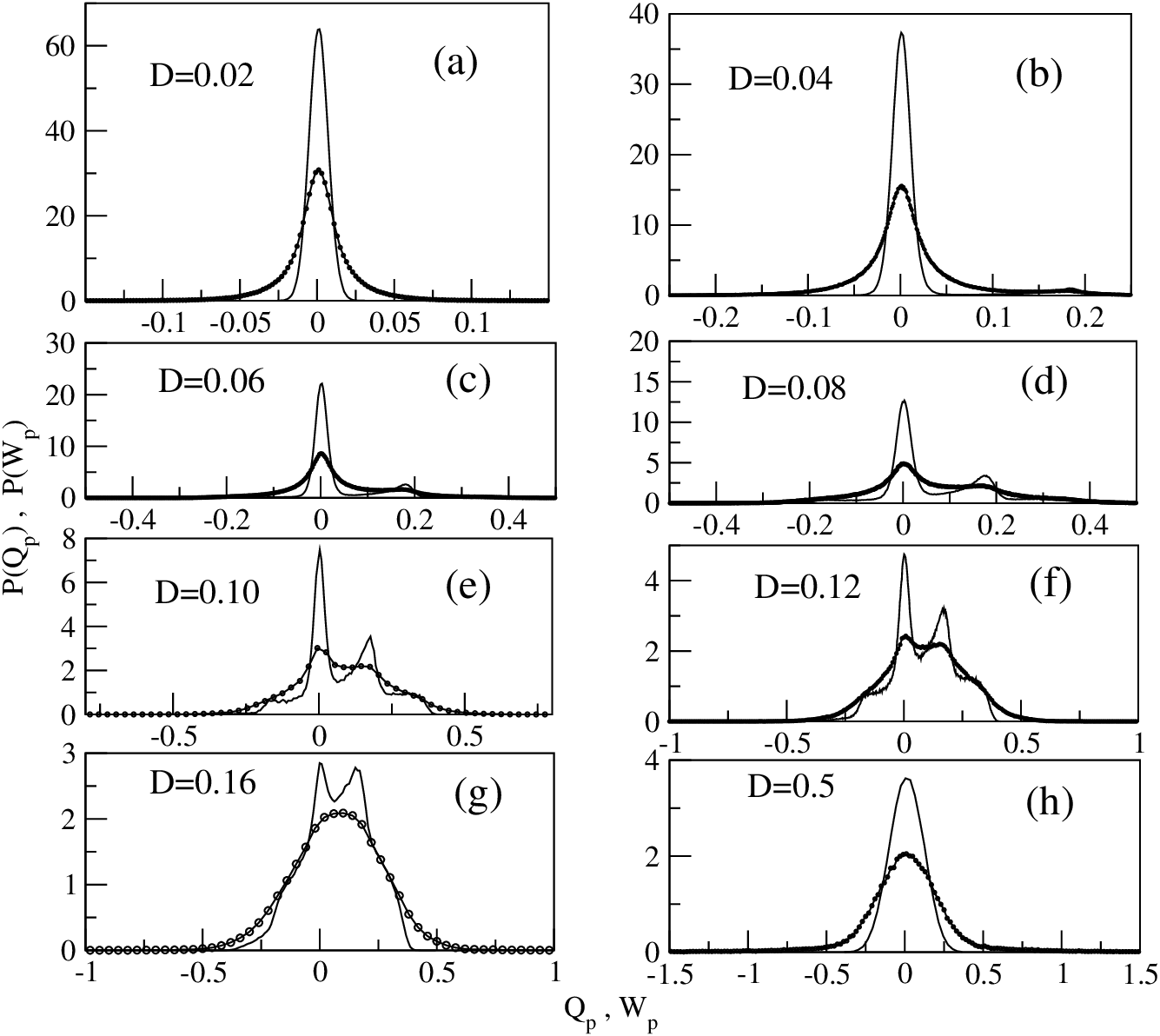}
\caption{} 
\end{center}
\end{figure}

\newpage
\begin{figure}[htp!]
\begin{center}
\input{epsf}
\includegraphics [width=6in,height=6.5in] {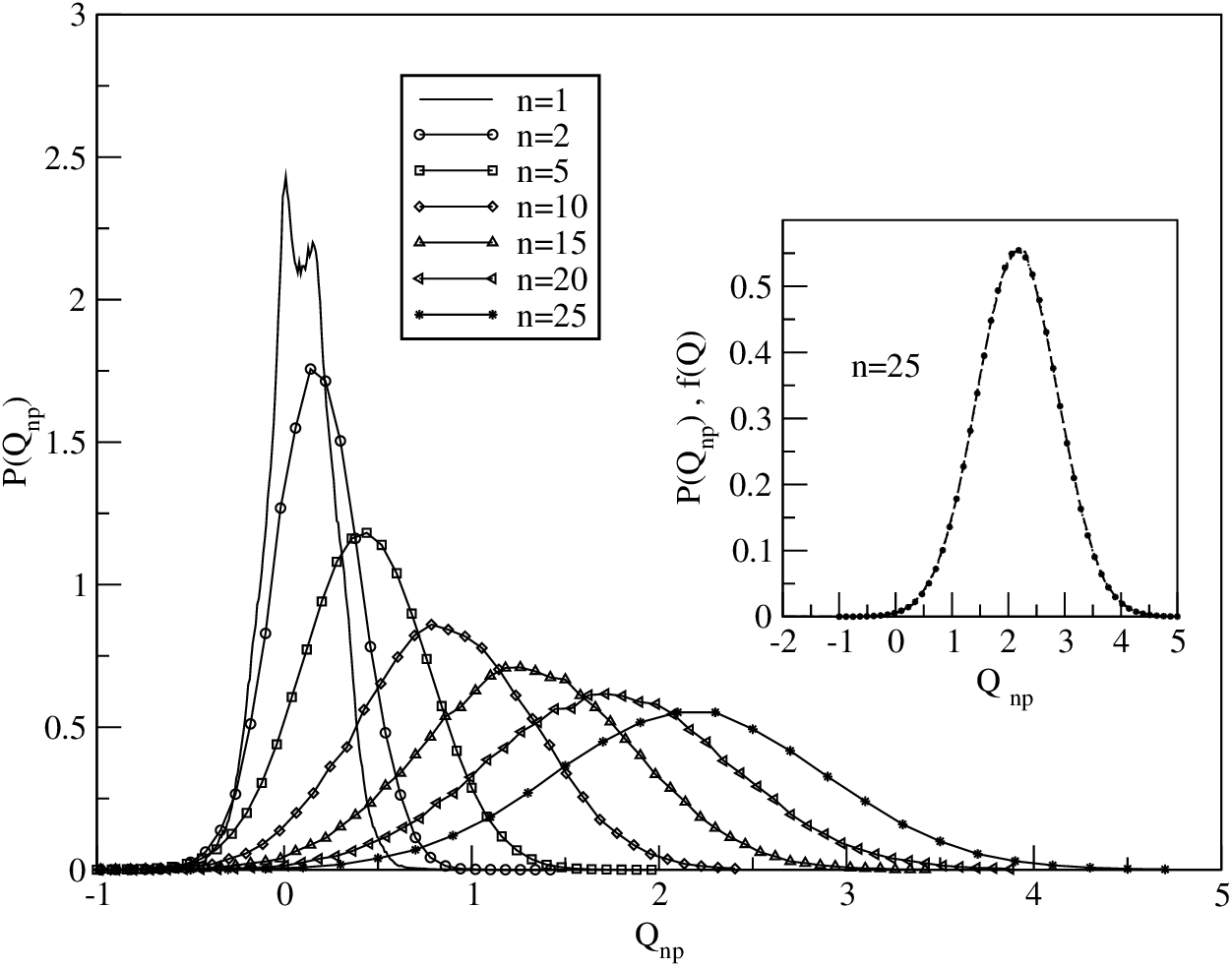}
\caption{} 
\end{center}
\end{figure}

\newpage
\begin{figure}[htp!]
\begin{center}
\input{epsf}
\includegraphics [width=6in,height=6.5in] {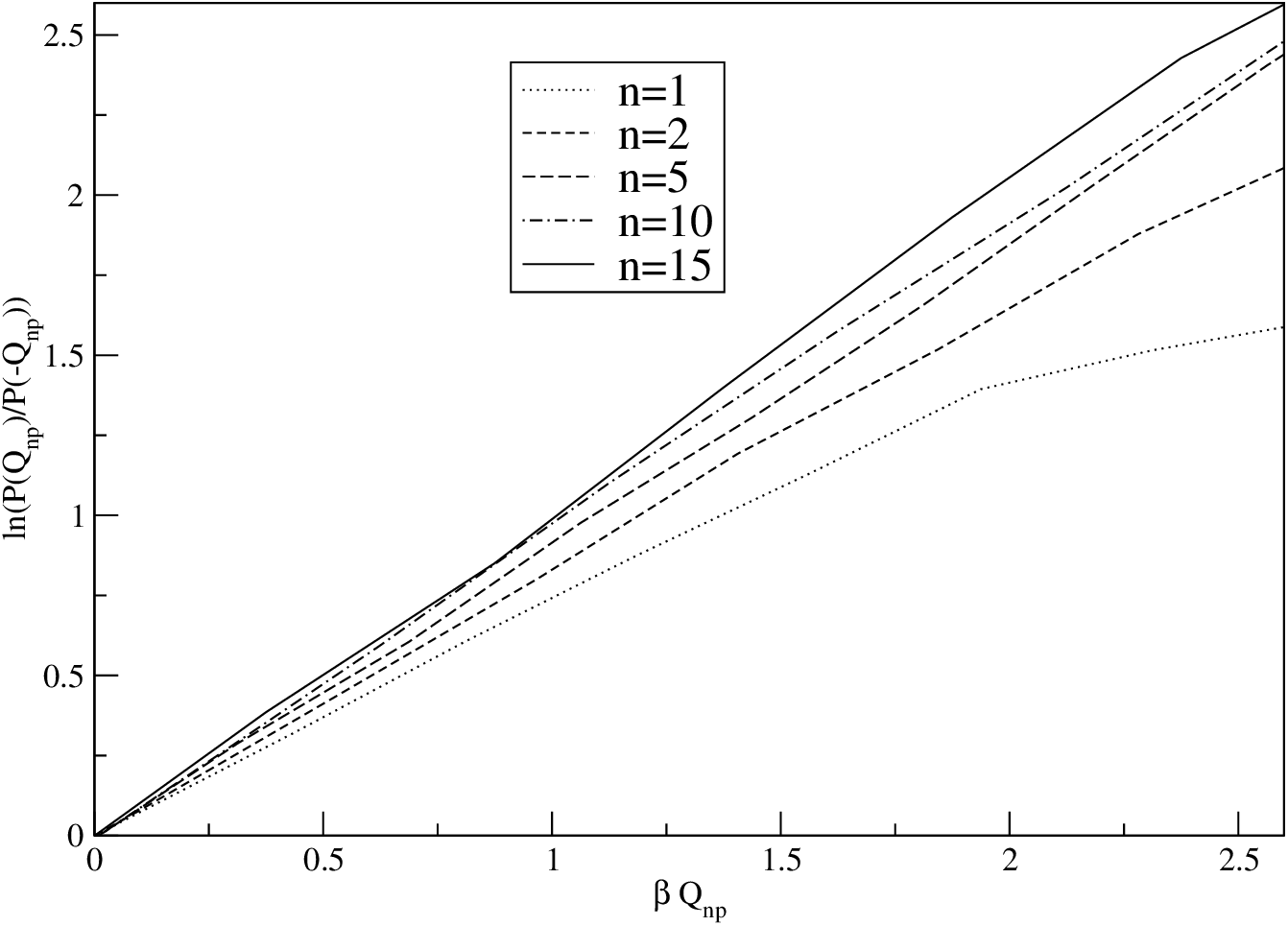}
\caption{} 
\end{center}
\end{figure}

\newpage
\begin{figure}[htp!]
\begin{center}
\input{epsf}
\includegraphics [width=6in,height=6in] {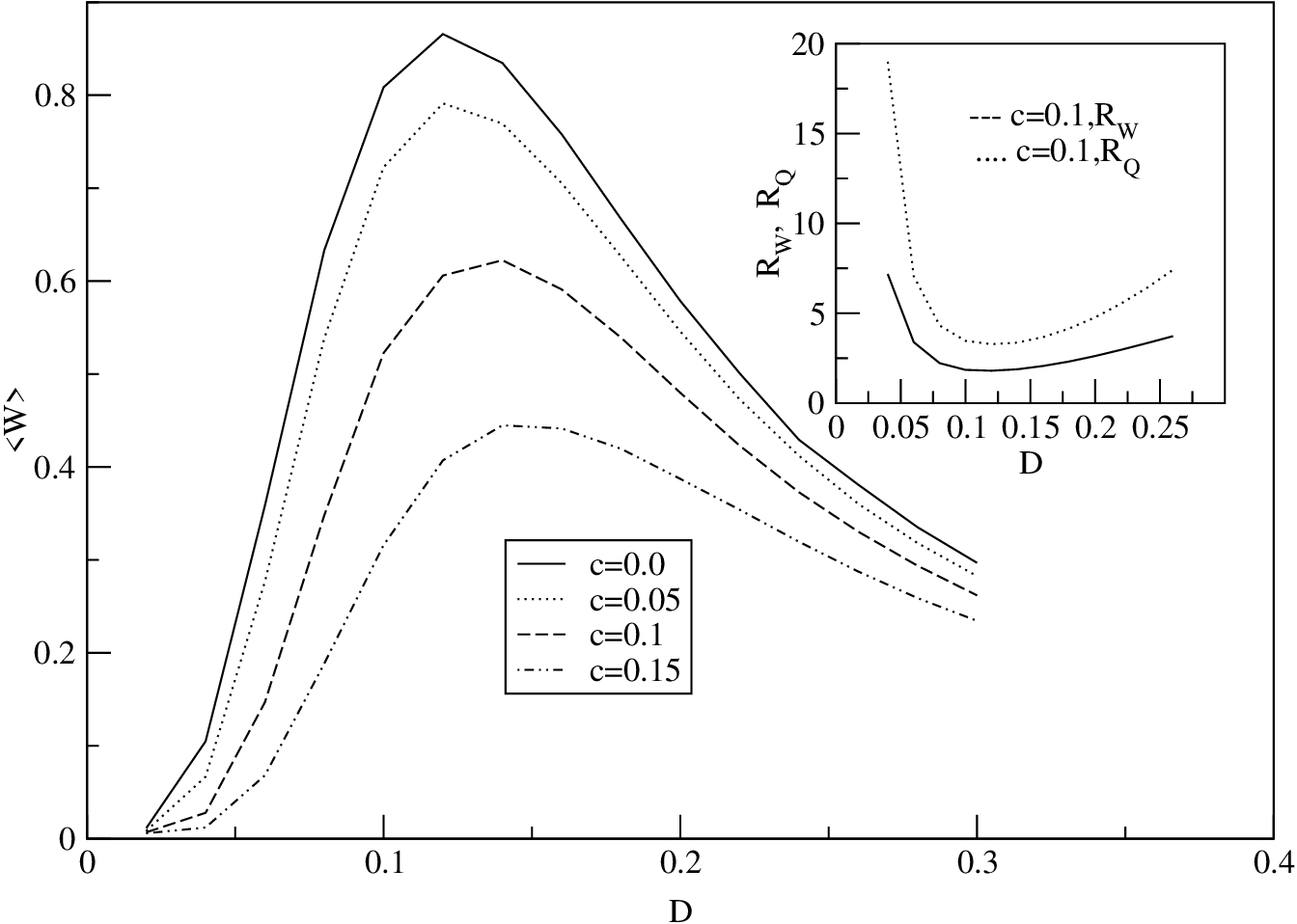}
\caption{} 
\end{center}
\end{figure}

\newpage
\begin{figure}[htp!]
\begin{center}
\input{epsf}
\includegraphics [width=6in,height=3.4in] {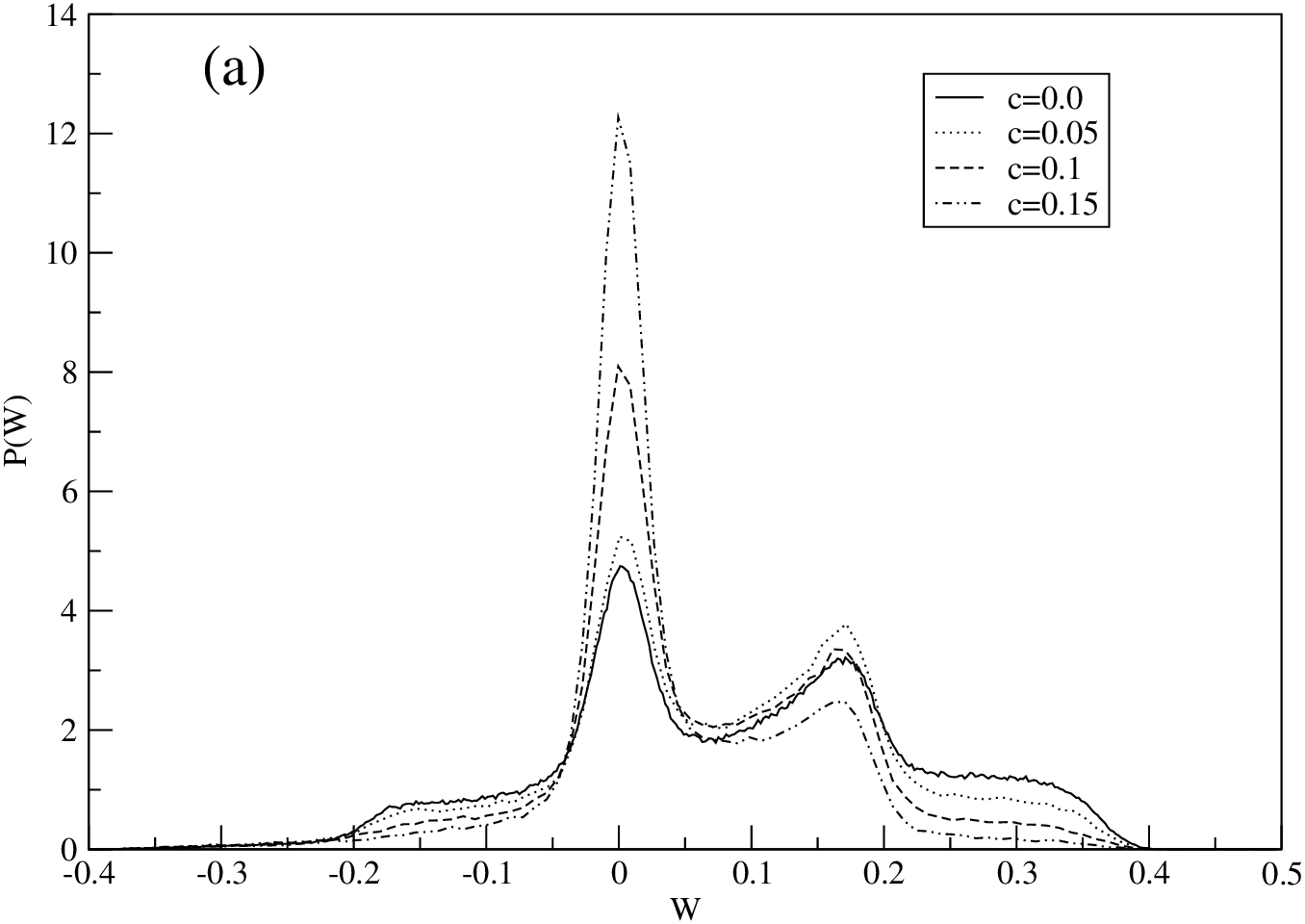}
\end{center}
\end{figure}
\begin{figure}[htp!]
\begin{center}
\includegraphics [width=6in,height=3.4in] {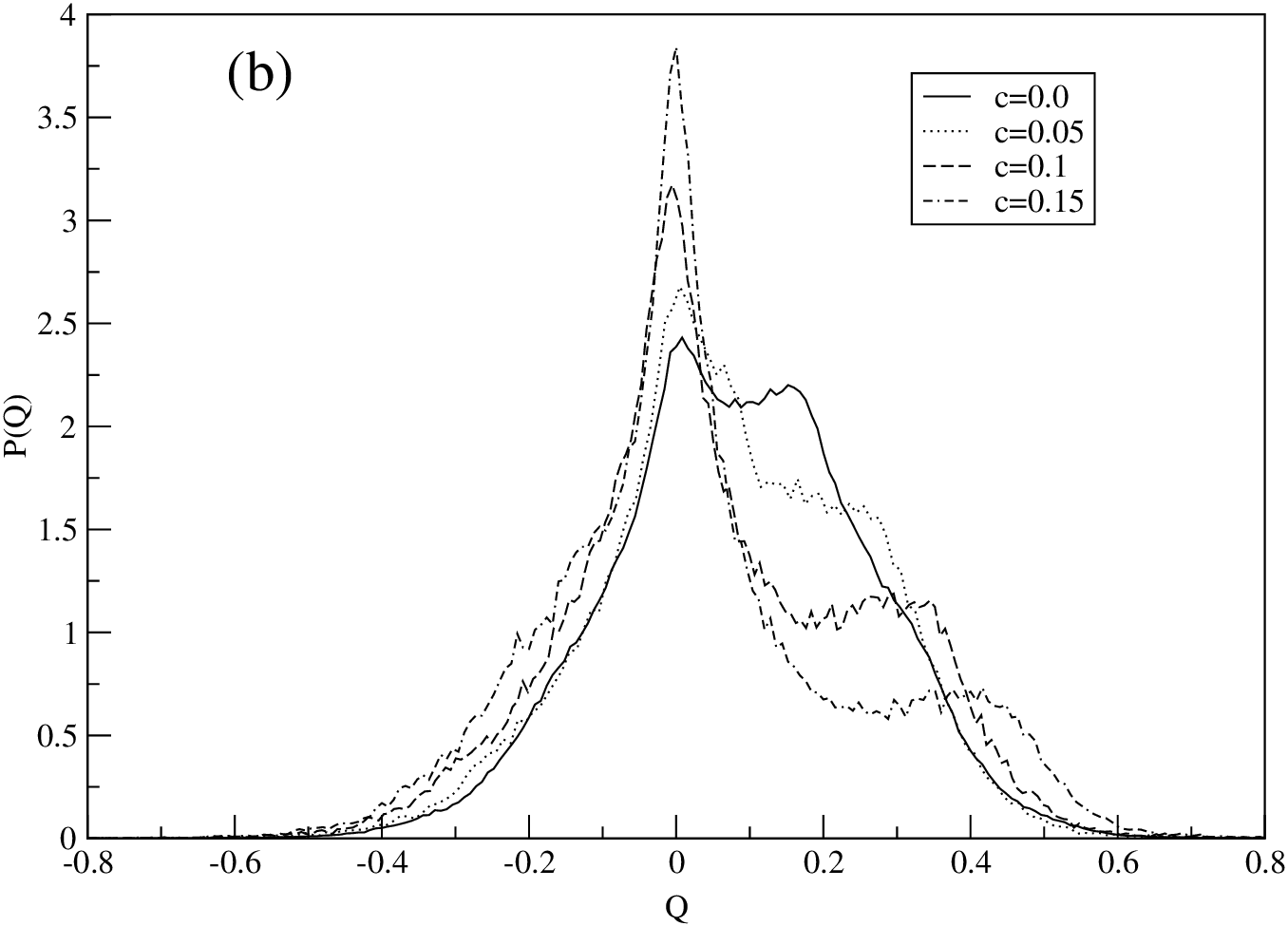}
\caption{} 
\end{center}
\end{figure}

\newpage
\begin{figure}[htp!]
\begin{center}
\input{epsf}
\includegraphics [width=6in,height=5in] {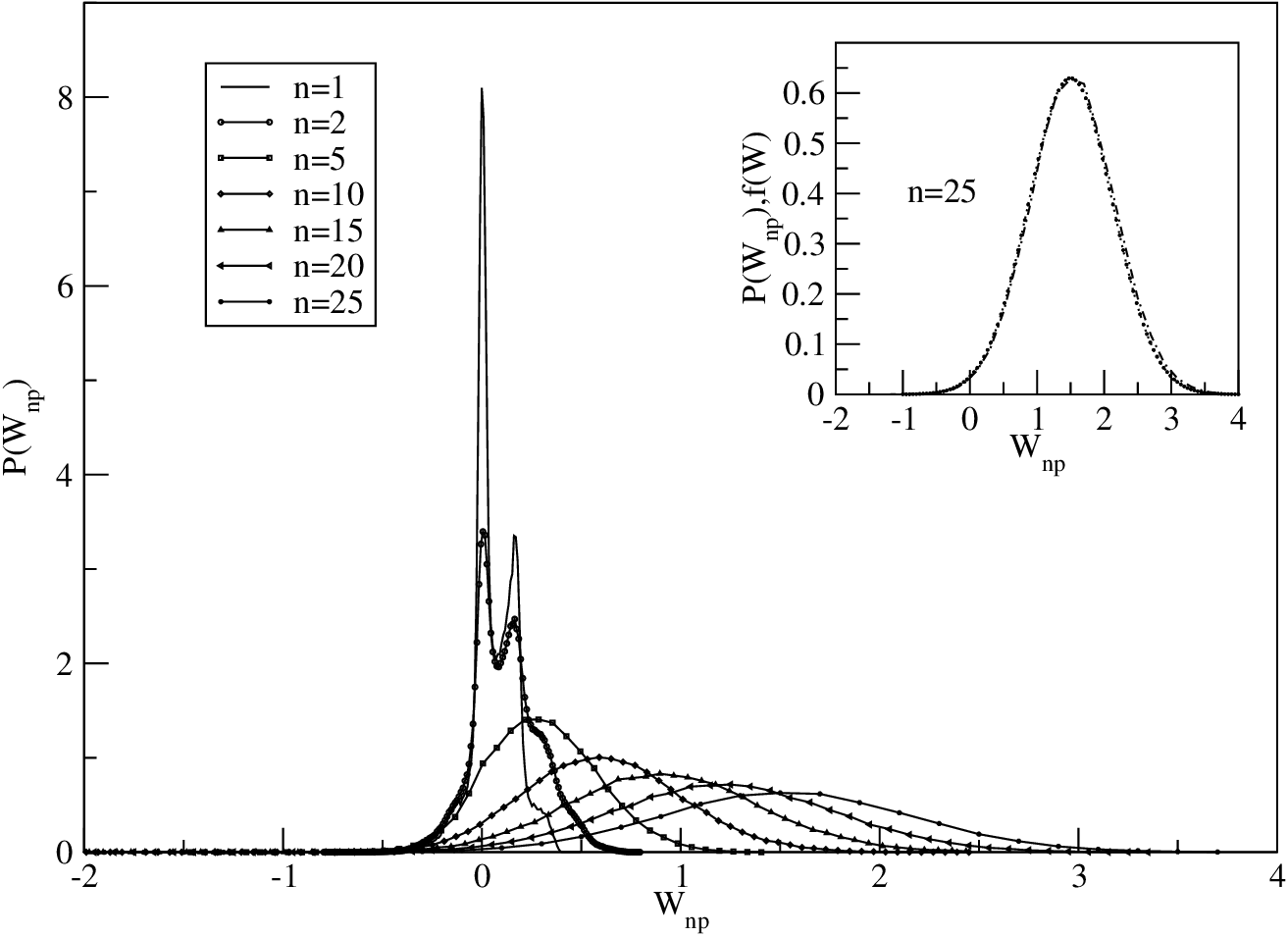}
\caption{} 
\end{center}
\end{figure}

\newpage
\begin{figure}[htp!]
\begin{center}
\input{epsf}
\includegraphics [width=6in,height=5in] {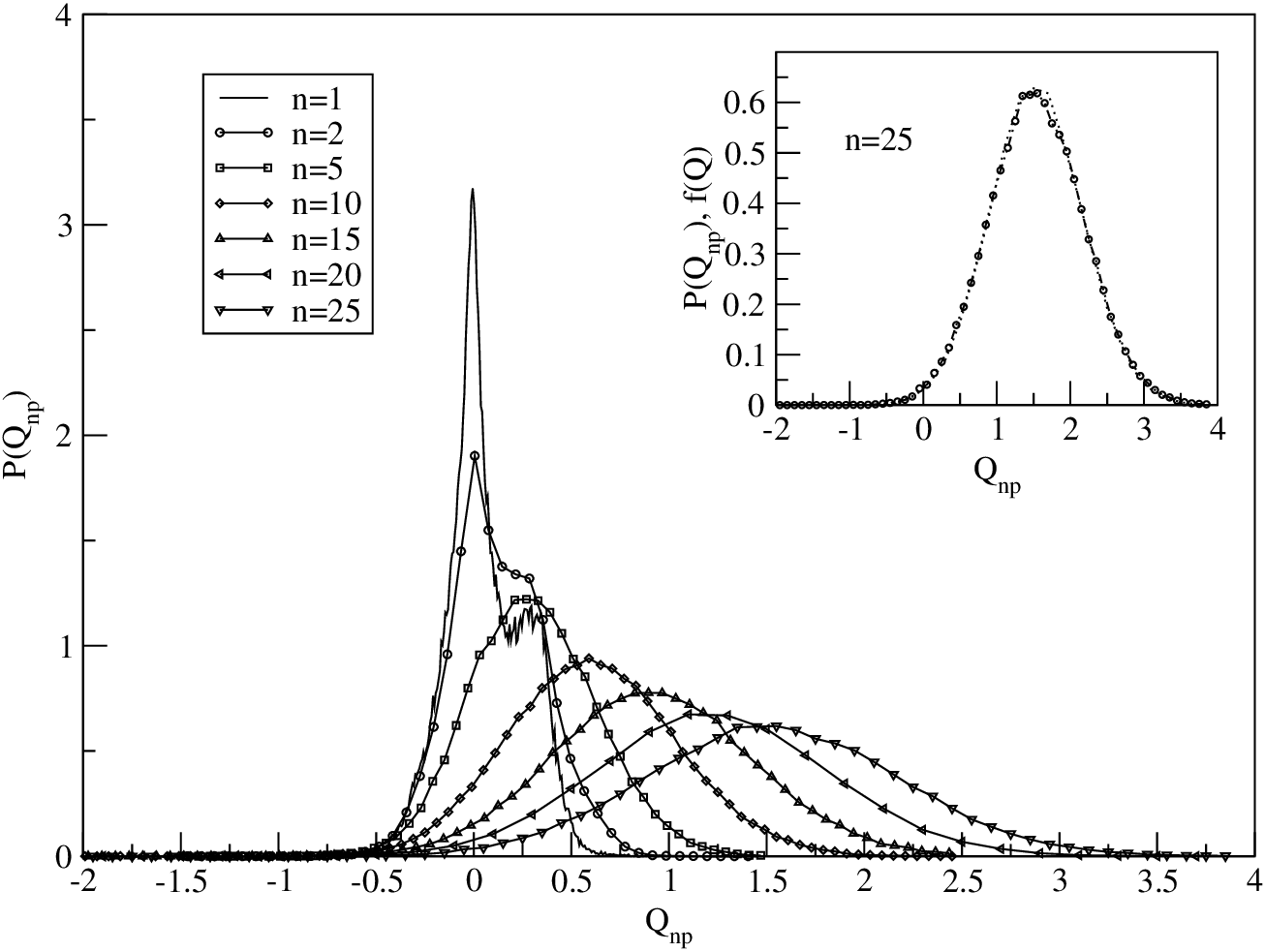}
\caption{} 
\end{center}
\end{figure}
\end{document}